\documentclass[a4paper,fleqn]{cas-sc}

\usepackage[authoryear,longnamesfirst]{natbib}

\def\tsc#1{\csdef{#1}{\textsc{\lowercase{#1}}\xspace}}
\tsc{WGM}
\tsc{QE}
\tsc{EP}
\tsc{PMS}
\tsc{BEC}
\tsc{DE}

\begin{document}
\let\WriteBookmarks\relax
\def\floatpagepagefraction{1}
\def\textpagefraction{.001}
\shorttitle{Microstructure evolution of carbon films}    
\shortauthors{Torres et al.}  
\title [mode = title]{Microstructure evolution of carbon films with increasing substrate temperature by using physical vapor deposition techniques} 
\author[1]{Sindry Torres}
[orcid=0000-0002-2003-2400]
\cormark[1]


\affiliation[1]{organization={Instituto de F\'{\i}sica Rosario},
            addressline={Bv. 27 de Febrero 210 bis}, 
            city={Rosario},
            postcode={S2000EZP}, 
            state={Santa Fe},
            country={Argentina}}
            
\affiliation[2]{organization={Facultad de Ciencias Exactas, Ingenier\'{\i}a y Agrimensura (UNR)},
            addressline={Av. Pellegrini 250}, 
            city={Rosario},
            postcode={S2000BTP}, 
            state={Santa Fe},
            country={Argentina}}
            
\affiliation[3]{organization={INTEMA, Facultad de Ingenier\'{\i}a, Universidad Nacional de Mar del Plata, Consejo Nacional de Investigaciones Cient\'{\i}ficas y T\'ecnicas (CONICET)},
            addressline={Av. Col\'on 10850}, 
            city={Mar del Plata},
            postcode={7600}, 
            state={Buenos Aires},
            country={Argentina}}

\affiliation[4]{organization={Instituto de Desarrollo Tecnol\'ogico para la Industria Qu\'{\i}mica, CONICET - UNL},
            addressline={G\"uemes 3450}, 
            city={Santa Fe},
            postcode={3000}, 
            state={Santa Fe},
            country={Argentina}}

\author[1]{ Giuliano Frattini}[orcid=0000-0003-2017-0106]
\author[3]{ Leonel I. Silva}[orcid= 0000-0003-3255-6882]

\author[1,2]{ Carlos E. Repetto}[orcid=0000-0003-3344-4997]
\author[4]{ Alejo Aguirre}[orcid=0000-0001-8797-8922]
\author[1,2]{ Bernardo J. G\'omez}[orcid=0000-0002-9573-2911]
\author[1,2]{ Ariel Dobry}[orcid=0000-0002-9487-538X]
\cortext[1]{Corresponding author}
\begin{abstract}
We study how the substrate temperature influences the structural properties of carbon films deposited by PVD (physical vapor deposition) techniques. We adapted a heating system inside the deposition chamber, with temperatures up to 700$^{\circ}$C. Here we develop an experimental setup that allows us to obtain large films of the desired material on any substrate, with deposition times of the order of a minute. The characterization is based mainly on the analysis of the Raman spectra, where the evolution of the {\bf{G}}  and {\bf{D}} peaks corresponding to the material in its amorphous phase is observed. With increasing substrate temperature, the $sp^{2}$ zones grow. A displacement to the right of the {\bf{G}} peak and an increase in the I(D)/I(G) ratio is seen. At 700$^{\circ}$C a 2D zone at a frequency greater than 2000 cm$^{-1}$ appears. Four Lorentzian-shaped bands are necessary to account for the peaks at this zone, whose centers correspond to different combinations of first-order ones. This indicates that we have a highly disordered sample and we are at the transition zone from amorphous carbon to a graphene layer. The Tauc gap energy ratio decreases as temperature increases indicating that there is a graphitization of the sample. 
Transmission FTIR study is carried out at some of the intermediate temperatures, determining the type of bond at low frequencies. These bonds are consistent with the ones of the hydrogenated amorphous carbon (a-C:H) structure.
\end{abstract}

\begin{highlights}
\item Development of an experimental setup to obtain large carbon nanofilms in short times by increasing the substrate temperature.

\item Evolution of the microstructure, as shown by Raman peaks position, from amorphous carbon to the first step of nanocrystalline structure. 
 
\item  Reduction of the optical gap with the substrate temperature indicating graphitization of the sample. 
\item Infrared active modes consistent with the vibration of hydrogenated amorphous carbon. 
\end{highlights}


\maketitle









\section{Introduction}
\label{Intro}
Graphene is an allotrope of carbon and has been widely studied in recent decades. It has a wide variety of properties and applications in different fields of science, from materials physics, nanoscience, and biological applications. There are several preparation methods for graphene, such as mechanical exfoliation of graphite, thermal reduction, laser ablation, and chemical vapor deposition (CVD) growth.  The most efficient method for preparing large graphene surfaces is chemical CVD, both for monolayer and multilayer growth as per \cite{Munoz2013,wang2020}. The CVD techniques have been generally used to make deposits on copper, or eventually another metal substrate, which after that, should be transferred to the desired materials. 

It would be desirable to develop techniques that allow depositing on other substrates directly without the transfer process.  

The present work is focused on the development of an experimental setup that allows to obtain sufficiently large graphene sheets on any substrate, using the PVD technique.
 Similar attempts have been made in previous work by \cite{chen2014optimization}. The deposition times in this work were of the order of hours. In the present work, we develop a PVD deposition system to go from amorphous carbon sheet to the first stages of graphene layers using deposition times of the order of a minute or less.

The paper is organized as follows:

In Section \ref{synthesis}, we describe the experimental setup, including the deposition techniques and the heating procedure to increase the substrate temperature.

In Section \ref{AFM}, we characterize the sample surface by AFM technique.

In Section \ref{sec:Raman}, we study the evolution of Raman spectra for samples deposited on different substrates as we increase the substrate temperature. We show that the higher the temperature, the larger the area with $sp^2$ bonds. At higher temperatures, we find Raman signal at high energy, in the so called 2D zone, which are double resonance peaks associated with the appearance of graphene pieces with defects.

In Section \ref{optical}, we study the optical properties both in the UV visible region as well as in the infrared region.

Finally, in Section \ref{conclusions}, we give the conclusions and perspectives.

\section{Synthesis of a-C:H films by an e-beam evaporator}
\label{synthesis}
The carbon films were deposited by the electron-beam physical vapor deposition technique, in which small graphite bars were bombarded with an electron beam that was generated in a tungsten filament powered by an external source at $150$ mA. This procedure was carried out in a high vacuum chamber at $10^{-8}$ Torr.
A heating system leading to temperatures up to the order of  $700^{\circ}$C was adapted inside the vacuum chamber. This system consists of an external circuit powered by a source that is connected in series to a voltage controller and in turn is connected in parallel to two heating resistors.

\begin{figure}[h!]
\centering
\includegraphics[angle=0,trim = 0cm 0cm 0cm 0.0cm, clip, width=8cm]{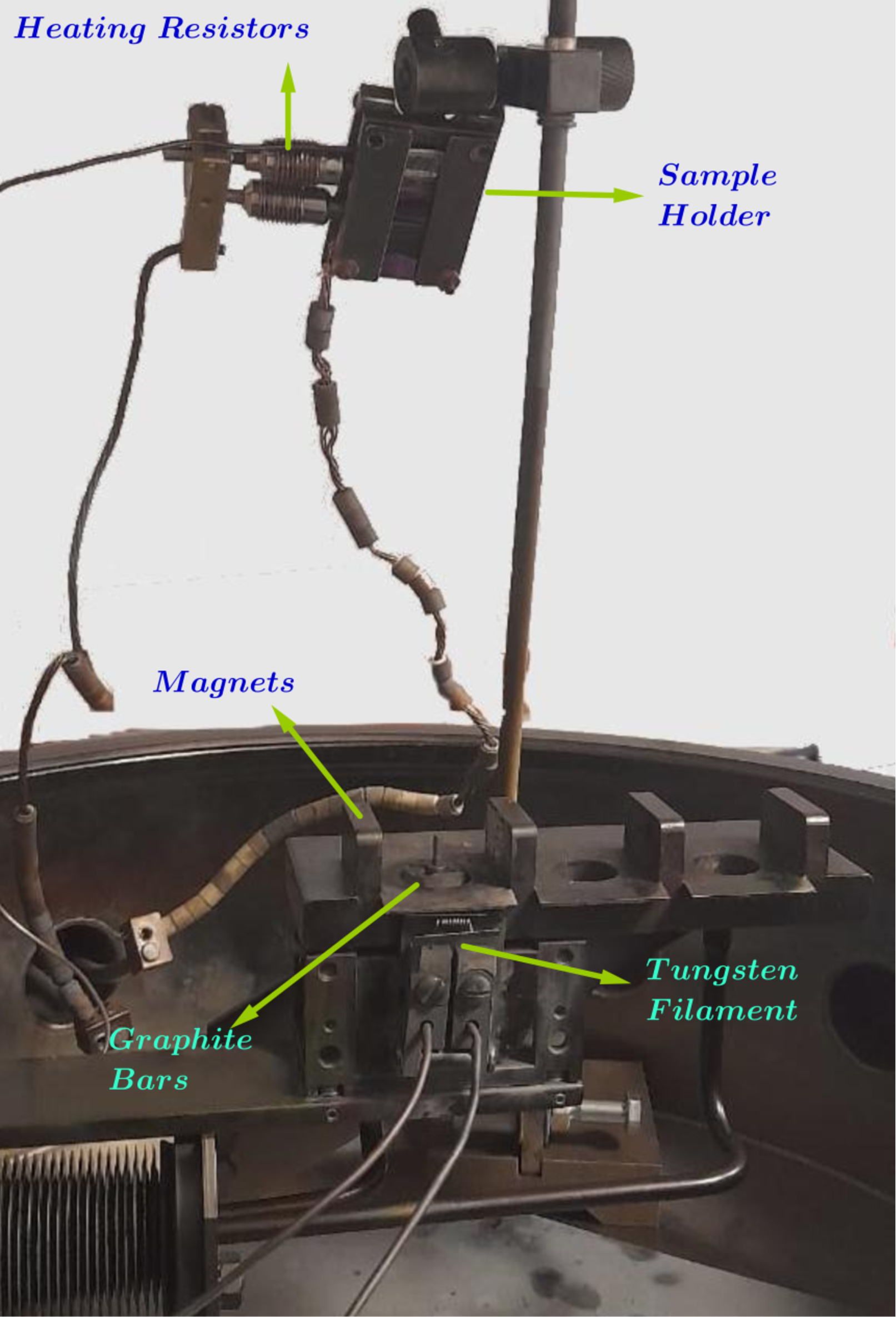}
\caption{Interior view of the vacuum chamber where amorphous carbon is deposited on different substrates. The temperature of the substrates can be raised using heating resistors.}
\label{fig:PVD}
\end{figure}

The position of the heating device is one of the essential factors to ensure the most homogeneous deposition process possible on the substrates. For all experiments, at the different temperatures and deposition times that will be shown in the next section, it remained in the same position as seen in Figure \ref{fig:PVD}. Vacuum conditions facilitate efficient material deposition by increasing the mean free path of the carbon vapor. The depositions were performed on glass substrates, silicon, and a $CaF_2$ window, starting at room temperature up to $700^{\circ}$C, with a deposition time of 30 seconds.

\begin{figure}[h!]
\centering
\vspace{0.6cm}
\includegraphics[angle=0,trim = 0cm 0cm 0cm 0.0cm, clip, width=15cm]{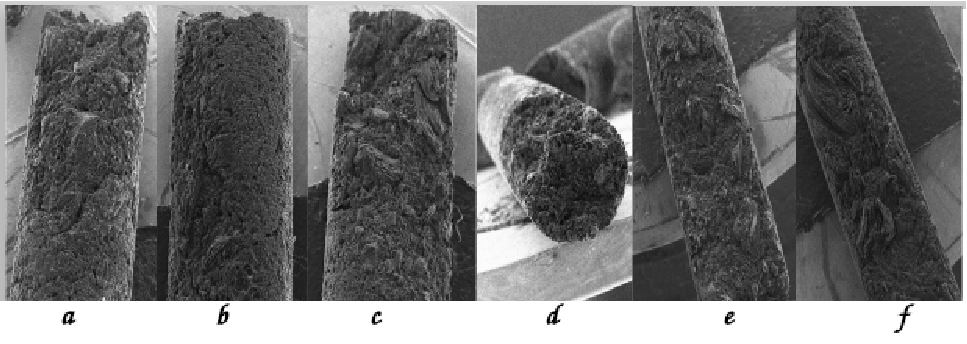}
\caption{The final effect on the graphite bars after evaporation of the carbon at room temperature (\textbf{\textit{a, b, c}}) and at  $600^{\circ}$C (\textbf{\textit{d, e, f}}).}
\label{fig:barras}
\end{figure}

In Section \ref{ramantimes}, we will show how the structure of the film changes when the incident time increases. Here we analyze the modification of the rods. The experiment was carried out on samples at room temperature and at $600^{\circ}$C with different incident times (30, 45, 60, and 90 s). We obtain the percentage of deposited material from the difference between the weight of the rod before and after the incidence of the electron beam. The final effect on the graphite bars after evaporation of the carbon is shown in Fig. \ref{fig:barras}. The graphite rods were consumed by the electron beam in the tip area of the rods, and the evaporated material corresponds to approximately 7 percent of the weight of the rods. Figure \ref{fig:barras} shows the scanning optical microscopy images labeled  \textbf{\textit{ a, b, c}} corresponding to the deposition at room temperature with times of (30s, 45s, 90s), respectively, and images \textbf{\textit{ d, e, f}} correspond to the deposition at $600^{\circ}$C with times of (30s, 45s, 60s), respectively. 

\section{Characterization of films by AFM} \label{AFM}
\begin{figure}[h!]
\centering
\includegraphics[angle=0,trim = 0cm 0cm 0cm 0.0cm, clip, width=12cm]{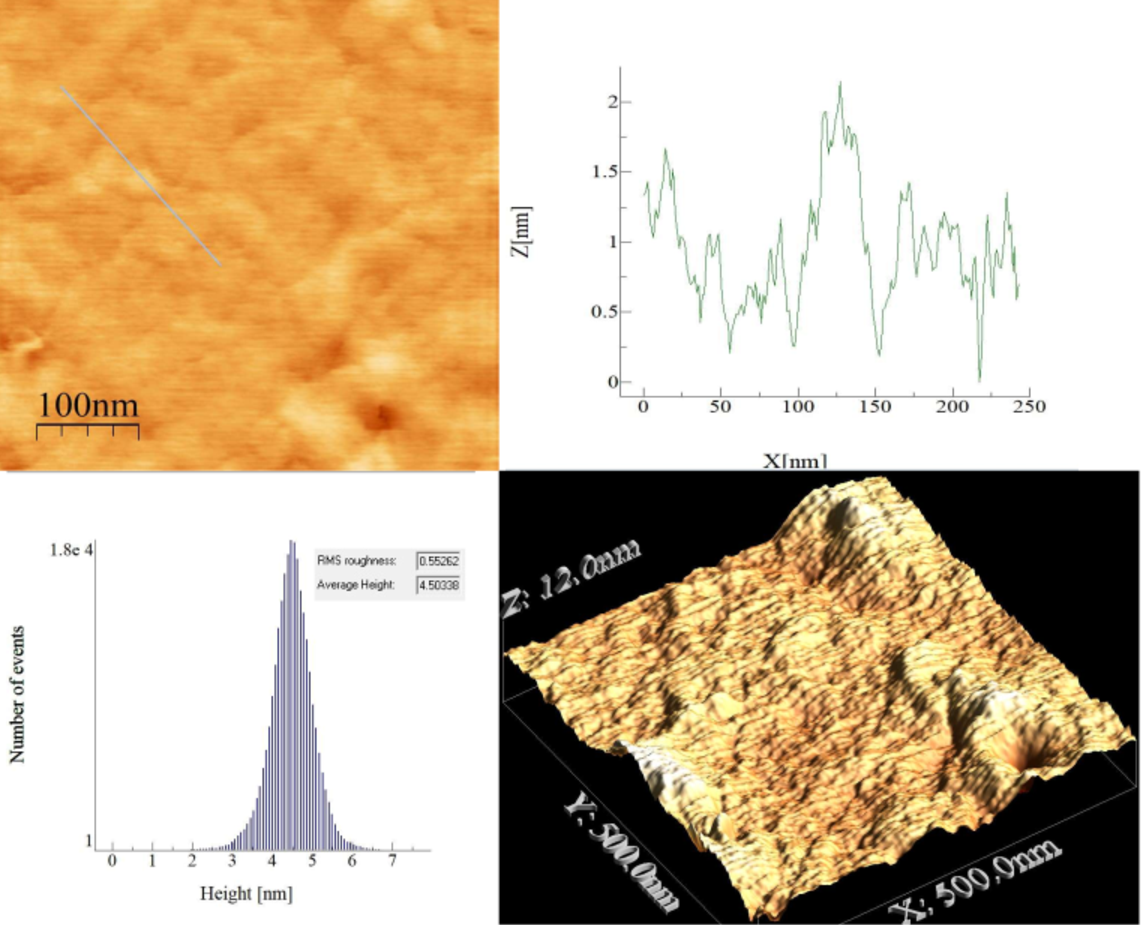}
\caption{Top left side: 2D image of the surface as seen by AFM. Top right side: depth mapping of the sample by traversing the path denoted in blue in the graph on the left. Bottom left side: histogram of the roughness on the same path. Bottom right side: 3D topographic image of the sample.}
\label{fig:afm}
\end{figure}

The topography of the deposited films was analyzed using a NanoTech Electronics atomic force microscope and WSxmM software for image processing. This was operated in the tapping mode with a silicon tip (150 KHz). The root mean squared (RMS) roughness value of 0.55 nm was obtained from the AFM (see the bottom left side of Fig. \ref{fig:afm}).

AFM measurements were performed on the carbon film deposited on silicon at  700$^{\circ}$C, showing a low roughness and fairly flat surface, following the substrate surface. It was scanned in the 500 nm range and a depth profile of the order of 2 nm can be observed. No periodic formations are observed on the entire surface, which shows that the samples are very thin (of the order of 30 nm).

%

\section{Evolution of Raman Spectra with the substrate temperature }
\label{sec:Raman}

The Raman spectrum shape fitting is a widely used method to study the detailed bonding structure of carbon films.
Therefore, we obtained the Raman spectra to characterize the structure of the deposited films. The spectra were acquired in a Renishaw In Via reflex system equipped with a charge-coupled device (CCD) detector of 1040$\times$256 pixels. A 514 nm diode laser (50 mW) was used as an excitation source in combination with a grating of 2400 grooves/mm and slit openings of 65 $\mu$m, which yield a spectral resolution of about 4 cm$^{-1}$. The laser power was kept below 10\% to avoid sample damage. A 50$\times$(0.5 NA) with a shorter working distance (210 $\mu$m) Leica metallurgical objective was used in the excitation and collection paths. Spectra were typically acquired in 10 s with at least 10 accumulations.

The sample had an area of 1 cm$^2$ and could be inhomogeneous. Therefore, Raman spectra were taken at different points. The spectra were fitted by taking into account different possibilities from Ref. \cite {Tai2009,Yuan2017}. Among the different alternatives, we chose to use  Gaussian functions for both peaks.

\begin{figure}[h!]
\begin{center}
\includegraphics[trim = 0cm 0cm 0cm 0.0cm, clip, width=26pc]{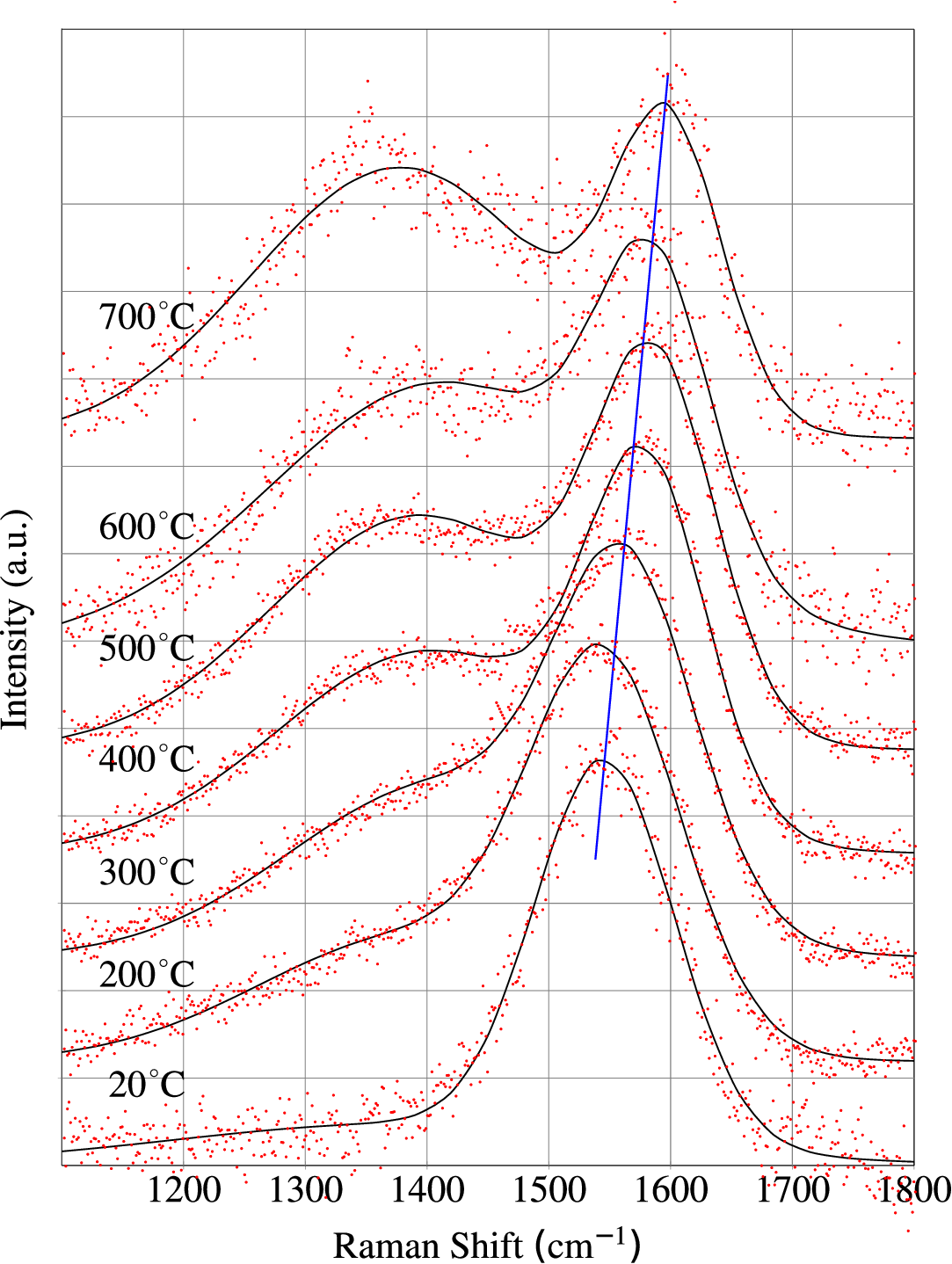}
\caption{In the Raman spectra for a-C:H silicon substrate, we observe the shift of the G peak as the temperature is increased for depositions on a silicon substrate.  The continuous line  is the superposition of the deconvolution of the peaks corresponding to the D and G modes. }
\label{fig:raman}
\end{center}
\end{figure}
 
Fig. \ref{fig:raman} shows the spectra of the samples deposited on silicon that were heated from room temperature to $700^{\circ}$C, with a deposition time of 30 seconds. It is worth mentioning that for samples deposited on glass, we obtained a quite similar Raman spectrum. 

\begin{figure}[h!]
\begin{center}
\includegraphics[trim = 0cm 0cm 0cm 0.0cm, clip, width=18pc]{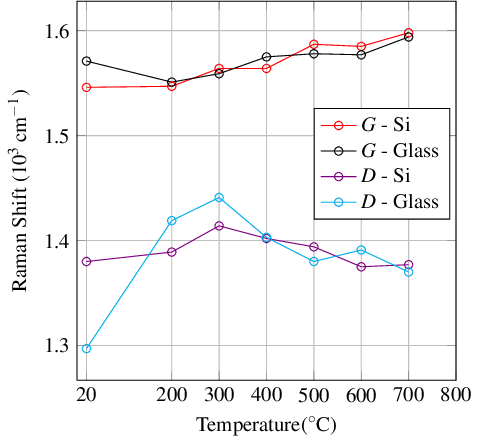}
\includegraphics[trim = 0cm 0cm 0cm 0.0cm, clip, width=18pc]{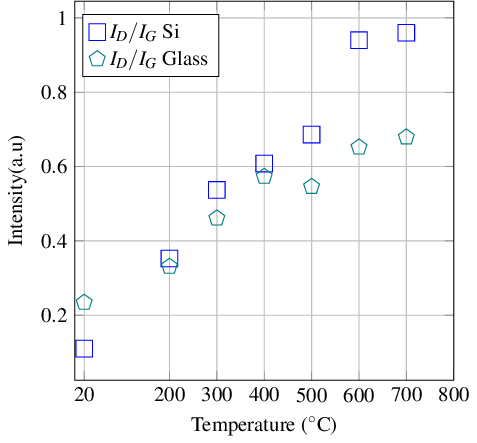}
\caption{On the right, it is shown the dependence of I$_D$/I$_G$ ratio on the deposition temperature of the films} 
\label{fig:posicion-intensidad}
\end{center}
\end{figure}

Once the background was subtracted, there were two main bands observed in all the spectra, corresponding to the D and G modes. They were located around 1400 and 1550 cm$^{-1}$ wavenumber, respectively. The D mode is related to the formation of  $sp^3$ bonds that represent the defects in the graphite structure. The G mode is associated with  $sp^2$ bond formation as per \cite{Robertson1993, ferrari2004raman}. The results on both substrates show that from room temperature, the structure of the films is completely amorphous.  With the increases of substrate temperature, the position of the G-peak and the relationship $I_D/I_G$  increase as shown in Fig. \ref{fig:posicion-intensidad}. This would imply that the size of the $sp^2$ zones has grown and the crystallization process has begun. \cite{ferrari2000interpretation}.    

Amorphous carbon structure consists of disordered hexagonal aromatic chains. By raising the deposition temperature, the formation of ordered rings without distortion will be increased. In this condition, the D peak intensity increases due to increasing the number of ordered hexagonal aromatic chains. Therefore, the $sp^2$ clusters increase, and a graphite-like structure will be formed by raising the deposition temperature. 
This conclusion is also based on the analysis of Ref. \cite{ferrari2000interpretation} 
which shows that for a-C the relation $I_D/I_G$ goes as $L_a^2$, being $L_a$ the typical size with $sp^2$ bonds. Note that for graphene or nano-graphite the expected relation changes as $I_D/I_G\propto 1/L_a$.  The relation between the intensity of the peaks decreases with the increase of $sp^2$ zones.
In Fig. 14 of \cite{ferrari2000interpretation}, it is shown that for a-C:H deposited by different techniques, the  G-peak shifts to the right and grows the $I_D/I_G$ when some $sp^3$ bonds drop. This means the $sp^2$ bonds have increased.

About the previous discussion, we remark that in the right part of Fig. 4, the relation $I_D/I_G$ in Si increases faster with the temperature. Therefore, in this comparison between substrates, silicon seems to favor the tendency for $sp^2$ bond formation and crystallization of the sample.  

\begin{figure}[!ht]
\begin{center}
\includegraphics[trim = 0cm 0cm 0cm 0.0cm, clip, width=24pc]{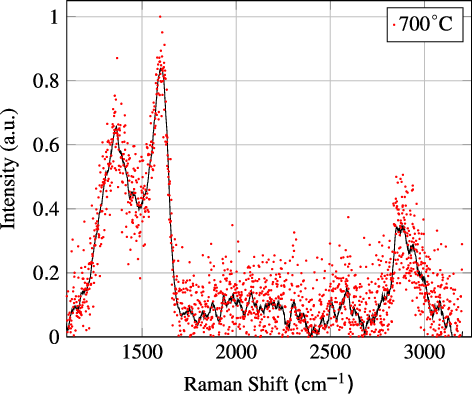}
\caption{Raman spectrum for samples at $700^{\circ}C$}
\label{fig:extendido-7}
\end{center}
\end{figure}

In Fig. \ref{fig:extendido-7}, we show the Raman spectra at  $700^{\circ}$C in a widely range of frequencies. The principal difference with the spectra at lower substrate temperatures is the appearance of a structure at a frequency greater than 2400 cm$^{-1}$. 

\begin{figure}[h!]
\begin{center}
\includegraphics[trim = 0cm 0cm 0cm 0.0cm, clip, width=25pc]{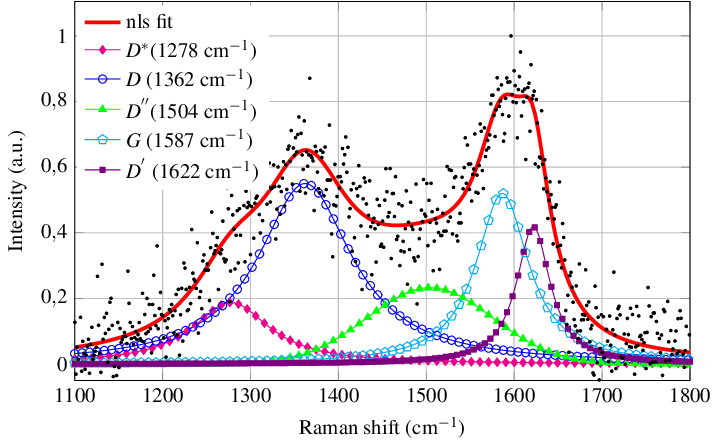}
\includegraphics[trim = 0cm 0cm 0cm 0.0cm, clip, width=25pc]{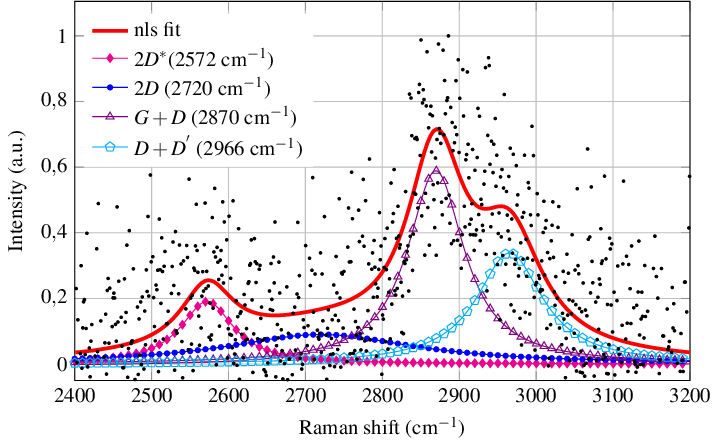}
\caption{(Top) First order region and (bottom) 2D region of the Raman spectra. In the right, we show the deconvolution with four Lorentzian. They correspond to 2$D^{*}$, 2$D$, $G+D$ (or $G+D^*$)  and $D+D^{'}$ modes. See text for explanations.}
\label{fig:raman-region2D}
\end{center}
\end{figure}

In Fig. \ref{fig:raman-region2D}, we split the first and second order spectra to identify the different modes. For the first order range, we identify three D peaks in addition to the D previously detected at lower temperatures.  For the assignation of the vibrational modes to each of these peaks see Table 1 of Ref. \cite{sadezky2005raman}.

The second order region was named by \cite{Wu2018} as a 2D region. The peaks in this region correspond to double resonance modes arising from combinations of the first-order peaks. By four-Lorentzian deconvolution and further fitting we can connect the peaks with the combinations given in the inset of the right part of  Fig. \ref{fig:raman-region2D}.
In particular, the 2D peak is characteristic of graphene and graphite. \cite{ferrari2013raman}. 

The emergence of other peaks in addition to the G and 2D indicates that the sample is yet disordered.\cite{ferrari2013raman}. We conclude that at this temperature the system is closer to a transition from amorphous carbon to a graphene layer. We expect that for higher temperatures graphene or multilayer graphene will start to develop. This will be signaled by a decrease of I$_D$/I$_G$ with further increasing the temperature and a shift of the G peak to the left in opposition to Fig. \ref{fig:raman}. 
This prediction is, in part, based on the results of Ref. \cite{chen2014growth} which show this behavior of the Raman peaks in the graphene samples.

\subsection{Study of the Raman spectra at different deposition times }\label{ramantimes}
\begin{figure}[h!]
\begin{center}
\includegraphics[trim = 0cm 0cm 0cm 0.0cm, clip, width=18pc]{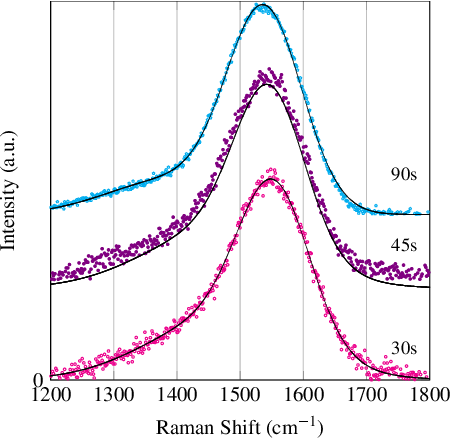}
\includegraphics[trim = 0cm 0cm 0cm 0.0cm, clip, width=18pc]{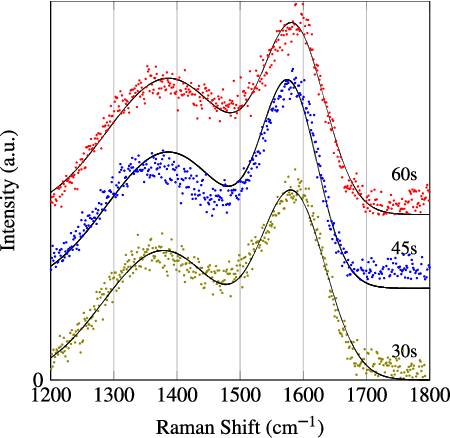}
\caption{The deconvolution of the G and D peaks at room temperature is shown on the left, and the deconvolution of the G and D peaks at $600^{\circ}C$ C is shown on the right, both for different deposition times.}
\label{fig:RamanTambas-sumadeconv}
\end{center}
\end{figure}

In the upper part of Fig. \ref{fig:RamanTambas-sumadeconv}, we show the deconvolution of the Raman spectra at different deposition times and two temperatures. They were deposited on a glass substrate. We would like to analyze if the deposition time changes the structure of the films. To answer this question we have studied the deconvolution of the Raman into the G and D peaks at room temperature for deposition times of 30, 45, 60, and 90 s. We have also studied the film deposited at 600$^{\circ}$C.   
In Fig. \ref{fig:RamanTambas-sumadeconv}, we show the sum of the deconvolution of the G and D peaks for different deposition times. At the bottom, we show the  $I_D/I_G$ for different deposition times.

In Table \ref{tableID_IGtiempo}, we show the position of the G and D peaks at different deposition times both at 20$^{\circ}$C and 600$^{\circ}$C.

In Fig. \ref{fig:RamanTambas-sumadeconv}, it is possible to see that at a given temperature the position and the structure of the peaks almost do not change with the deposition times. A possible explanation is that the heat transfer between the material that is being deposited and the substrate is a slower process than the one that takes place between the new material deposited and the one already present.

\begin{table}[!ht]
\begin{center}
\begin{tabular}{||c | c | c | c | c | c | c ||}  \hline
 Time[s]&$x_D$(20$^{\circ}C$)& $x_G$ (20$^{\circ}C$) & $I_D/I_G$ (20$^{\circ}C$)& $x_D$(600$^{\circ}C$)& $x_G$(600$^{\circ}C$) & $I_D/I_G$ (600$^{\circ}C$)\\ [0.5ex] 
 \hline\hline
  $30$  &1474 & 1554 &0.406 & 1378 & 1584 & 0.731   \\ \hline
  $45$  &1470 & 1547 &0.380 & 1372 & 1579 & 0.613  \\ \hline
  $60$  &--- & --- &--- & 1386 & 1586 & 0.782   \\ \hline
  $90$  &1391 & 1539 & 0.176& ---   & --- & --- \\ \hline
  \end{tabular}
\caption{Position of the D and G peaks and $I_D/I_G$ relation. Results for different deposition times and two temperatures of the substrate are shown.}
  \label{tableID_IGtiempo}
\end{center}
\end{table}

The positions of G and D peaks hardly change concerning each other, and the ratio of intensities fluctuates very little. It has been determined that the ratio of intensities remains unchanged over time.

\section{Optical Properties}\label{optical}
\subsection{UV Spectroscopy- Tauc Gap}

Data acquisition was performed on the JASCO model V-530 double beam UV VIS spectrophotometer with a spectral range of 200-1100 nm.
In the case of the amorphous carbon films studied in this work, they can be treated as a semiconductor. Usually, the energy gap $E_g$ in crystalline and amorphous semiconductors is determined by the Tauc method as described by  \cite{salwank1993optical}, i.e. linear extrapolation of the plot of $(\alpha h \nu)^{1/m}$ versus the photon energy, $h\nu$, 
\begin{equation}
    \left(\alpha h \nu\right)^{1/m} = a\left(h\nu - E_g \right)\,,
\label{ec.Tauc}
\end{equation}
where $m$ is a constant that accounts for the type of optical transition. Since we do not know the nature of the optical transition, to determine the energy gap, we follow the method described by  \cite{Jarosinski2019inverse}, which is based on transforming the measurement data by using an inverse logarithmic derivative (ILD) and performing linear ﬁttings. In this way, the value of the energy gap and the parameter $m$ can be obtained simultaneously, 
\begin{equation}
    \frac{\Delta(h\nu)}{\Delta \ln(\alpha h \nu)} = \frac{1}{m}(h\nu-E_g)\,.
\label{ec.ILD}
\end{equation}
Another advantage of this method is that it can be carried out without knowing the thickness of the samples.

\begin{figure}[h!]
\begin{center}
\includegraphics[angle=0,trim = 0cm 0cm 0cm 0.0cm, clip, width=12cm]{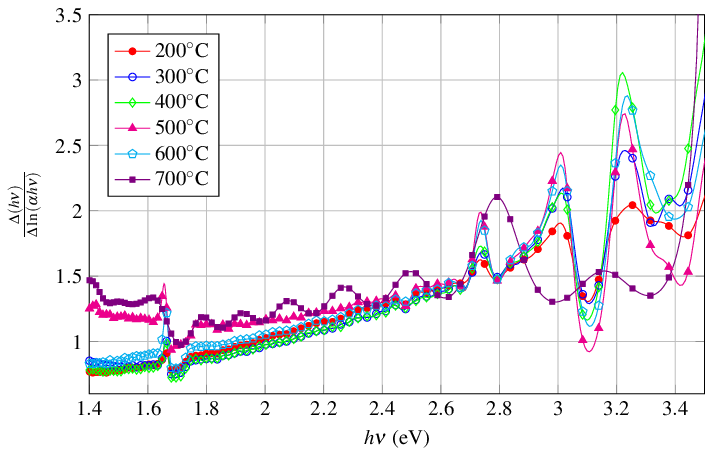}
\caption{Inverse logarithmic derivative method performed on data obtained from the visible UV spectrum for films deposited on glass substrates at different temperatures, with a deposition time of 30 seconds.}
\label{fig:derivadas}
\end{center}
\end{figure}   
Then, starting from the data obtained from the visible UV spectrum for films deposited on glass substrates at different temperatures, with a deposition time of 30 seconds, we calculated the ILD and plotted it as a function of $h\nu$, as can be seen in Fig. \ref{fig:derivadas}. In the region where the behavior is linear, we fit with a straight line from which we determine the exponent $1/m$ and the gap $E_g$, as it was described in Equation (\ref{ec.ILD}), from the slope and the abscissa to the origin, respectively. The values obtained are shown in columns 2 and 3 of Table 4.

\begin{figure}[h!]
\begin{center}
\includegraphics[angle=0,trim = 0cm 0cm 0cm 0.0cm, clip, width=12cm]{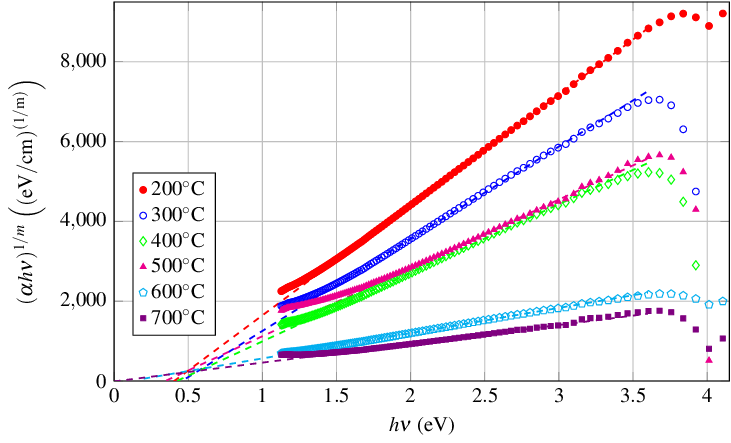}
\caption{The dashed lines are linear fitting of part of the spectra as has been done from the pioneering work of \cite{tauc1968optical}. $E_g$ is obtained by extrapolation (see text).}
\label{fig:Tauc}
\end{center}
\end{figure}

Taking into account the values of the exponent determined in this way, to check the consistency of the method, we proceeded to find the gap with the traditional Tauc method, following Equation (\ref{ec.Tauc}). The behavior of $(\alpha h \nu)^{1/m}$ vs. $h\nu$ for the different temperatures is shown in Fig. \ref{fig:Tauc}. The results obtained for the gap can be seen in column 4 of Table 4. It can be observed that the values are very close to those obtained by the ILD method and that the uncertainties are smaller.

\begin{table}[h!]
\begin{center}
\begin{tabular}{||c | c | c | c ||}  
\hline
Temperature ($^{\circ}$C) & $1/m$ ILD & $E_g$ ILD (eV) & $E_g$ Tauc (eV)\\ [0.5ex] 
\hline\hline
200 & 0.641 $\pm$ 0.004 & 0.41 $\pm$ 0.01 & 0.405 $\pm$ 0.001 \\ \hline
300 & 0.644 $\pm$ 0.004 & 0.49 $\pm$ 0.01 & 0.457 $\pm$ 0.002  \\ \hline
400 & 0.631 $\pm$ 0.004 & 0.460 $\pm$ 0.009 & 0.423 $\pm$ 0.002   \\ \hline
500 & 0.64 $\pm$ 0.03 & 0.3 $\pm$ 0.1 & 0.346 $\pm$ 0.004  \\ \hline
600 & 0.570 $\pm$ 0.004 & 0.15 $\pm$ 0.01 & 0.108 $\pm$ 0.003  \\ \hline
700 & 0.60 $\pm$ 0.03 & 0.01 $\pm$ 0.01 & 0.024 $\pm$ 0.003  \\ \hline
\end{tabular}\label{Tabla4}
\caption{ Values of band gap for different temperatures and values from exponent $n$ equation Tauc.}
\end{center}
\end{table}

From the values shown in this Table, we see that while the $1/m$ exponent maintains a fairly stable value, the gap remains almost constant up to 400$^\circ$C and decreases appreciably as the temperature further increases, approaching zero at 700$^\circ$C. Note that this value is close to $m=3/2$, corresponding to direct forgiven transitions.

This result is consistent with the one obtained in section \ref{sec:Raman} when we approach 700$^\circ$C.
The optical properties of amorphous carbon are directly related to the ratio of the $sp^2$ and $sp^3$ bonds \cite{Robertson1993}.
In the samples at higher temperatures, the optical band gap with very small values close to zero and the Raman results indicate that the samples tend to have an order, which means that the $sp^3$ bonds are decreasing with an increase in $sp^2$ bonds suggesting that a graphitization is taking place in the sample or, in other words, that the sample is moving from disorder to order.

\subsection{Infrared Spectroscopy}
For the transmission FTIR measurements, the samples were placed into a Pyrex IR cell fitted with $CaF_2$ windows. Infrared spectra (25 scans) were recorded at a resolution of 4 cm$^{-1}$ using an FTIR spectrometer (Thermo iS50 with a cryogenic MCT detector). The $CO_2$ and water vapor contributions from the atmosphere to the spectra were eliminated by purging continuously the bench of the spectrometer and the optical path with purified air (Parker Balston FTIR purge gas generator).
\begin{figure}[h!]
\includegraphics[angle=0,trim = 0cm 0cm 0cm 0.0cm, clip, width=12cm]{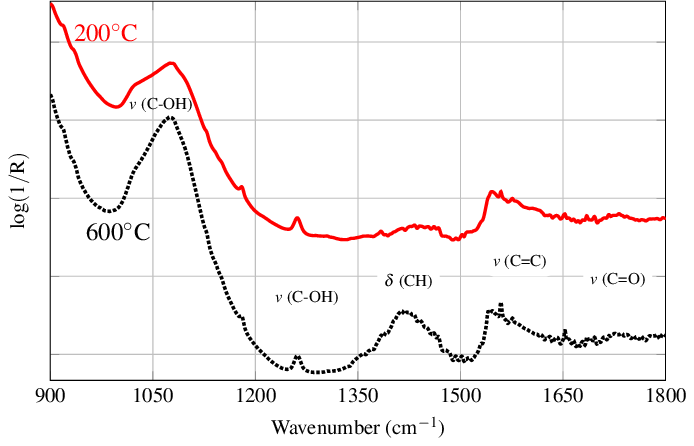}
\caption{The FTIR spectrum in low frequency region.}
\label{fig:CAF2-ALL}
\end{figure}
The FTIR study is carried out at some of the intermediate temperatures: 200$^{\circ}$C and 600$^{\circ}$C determining the type of bond at low frequencies. Fig. \ref{fig:CAF2-ALL} shows the spectra of the depositions performed at different temperatures together with the assignment of the main absorption bands. The spectrum shows a close band in the region 1730 cm$^{-1}$ at both temperatures, corresponding to the vibrational modes [$\nu$(C=O)] of the carbonyl group, a band appears near 1550 cm$^{-1}$ assigned to the vibrational modes [$\nu$(C=C)]  belonging to the formation of the $sp^2$ bond, these vibrational modes are of the stretching symmetrical type. Wide adsorption appears in the 1460 cm$^{-1}$ range, which represents the combined contribution of the [$\delta$(CH)] and the  [$\nu$(C=C)] vibrational modes, due to the increase in the temperature as seen in the 600$^{\circ}$C, this mode is called the Bending symmetrical type. Near 1260 cm$^{-1}$ and 1078 cm$^{-1}$ a minor shift of both spectra is noticed, identifying a band  [$\nu$(C-OH)] corresponding to the oxhydril group and Bending symmetrical type. Finally, this spectrum could be associated with the IR active low-frequency vibration modes of hydrogenated amorphous carbon.

\section{Conclusions}\label{conclusions}
We have developed a heating method to increase the substrate temperature in a PVD camera.
By evaporating a graphite bar and heating the substrate we obtained samples of carbon nanofilm and studied the evolution of the microstructure. By different experimental techniques, we have characterized the film. We found that the higher the temperature, the greater the areas of the film with sp$^2$ bonds. This fact shows that the system tends to go from amorphous carbon to multilayer graphene. Even though at 700$^{\circ}$C we found signal characteristics of graphene with defects. We conclude that at this temperature, we are close to a transition from amorphous to crystalline graphene. 

Our method opens up the possibility of obtaining large surface area carbon films on virtually any substrate. Moreover, this process could be carried out in times of the order of a few seconds.

Finally, in previous works, we, \cite{frattini2021effect, epeloa2018resistivity}, have shown that a-C:H could be used as resistivity humidity sensors. In future works, we will study the transport properties and the sensitivity to the relative humidity of the surrounding ambient, of the samples deposited at higher temperatures.  

\section{Acknowledgment}
This work was supported by CONICET through PUE-IFIR  0076, and Universidad Nacional de Rosario, through 80020190300213UR. Leonel Silva acknowledges Agencia Nacional de Promoci\'on Cient\'ifica y Tecnológica, Fondo para la Investigaci\'on Cient\'ifica y Tecnológica (PICT16‐3633) for financial support.

\bibliographystyle{model1-num-names}
\bibliography{torrestext}

\begin{thebibliography}{17}
\expandafter\ifx\csname natexlab\endcsname\relax\def\natexlab#1{#1}\fi
\providecommand{\url}[1]{\texttt{#1}}
\providecommand{\href}[2]{#2}
\providecommand{\path}[1]{#1}
\providecommand{\DOIprefix}{doi:}
\providecommand{\ArXivprefix}{arXiv:}
\providecommand{\URLprefix}{URL: }
\providecommand{\Pubmedprefix}{pmid:}
\providecommand{\doi}[1]{\href{http://dx.doi.org/#1}{\path{#1}}}
\providecommand{\Pubmed}[1]{\href{pmid:#1}{\path{#1}}}
\providecommand{\bibinfo}[2]{#2}
\ifx\xfnm\relax \def\xfnm[#1]{\unskip,\space#1}\fi
\bibitem[{Al-Ani(1993)}]{salwank1993optical}
\bibinfo{author}{Al-Ani, S.K.J.}, \bibinfo{year}{1993}.
\newblock \bibinfo{title}{Determination of the optical gap of amorphous
  materials}.
\newblock \bibinfo{journal}{Journal of Electronics} \bibinfo{volume}{75},
  \bibinfo{pages}{1153--1163}.
\bibitem[{Chen et~al.(2014a)Chen, Fan, Wang, Pan, Chen, Xu, Zou and
  Wu}]{chen2014optimization}
\bibinfo{author}{Chen, S.}, \bibinfo{author}{Fan, L.}, \bibinfo{author}{Wang,
  J.}, \bibinfo{author}{Pan, Y.}, \bibinfo{author}{Chen, F.},
  \bibinfo{author}{Xu, P.}, \bibinfo{author}{Zou, C.}, \bibinfo{author}{Wu,
  Z.}, \bibinfo{year}{2014}a.
\newblock \bibinfo{title}{The optimization of a self-focusing e-beam evaporator
  for carbon evaporation and the application for graphene growth}.
\newblock \bibinfo{journal}{Surface and Coatings Technology}
  \bibinfo{volume}{258}, \bibinfo{pages}{1196--1201}.
\bibitem[{Chen et~al.(2014b)Chen, Ma, Zhu, Yue, Hu, Chen and
  Wang}]{chen2014growth}
\bibinfo{author}{Chen, Y.N.}, \bibinfo{author}{Ma, T.B.}, \bibinfo{author}{Zhu,
  P.Z.}, \bibinfo{author}{Yue, D.C.}, \bibinfo{author}{Hu, Y.Z.},
  \bibinfo{author}{Chen, Z.}, \bibinfo{author}{Wang, H.},
  \bibinfo{year}{2014}b.
\newblock \bibinfo{title}{Growth mechanism of hydrogenated amorphous carbon
  films: Molecular dynamics simulations}.
\newblock \bibinfo{journal}{Surface and Coatings Technology}
  \bibinfo{volume}{258}, \bibinfo{pages}{901--907}.
\bibitem[{Epeloa et~al.(2019)Epeloa, Repetto, G{\'o}mez, Nachez and
  Dobry}]{epeloa2018resistivity}
\bibinfo{author}{Epeloa, J.}, \bibinfo{author}{Repetto, C.E.},
  \bibinfo{author}{G{\'o}mez, B.J.}, \bibinfo{author}{Nachez, L.},
  \bibinfo{author}{Dobry, A.}, \bibinfo{year}{2019}.
\newblock \bibinfo{title}{Resistivity humidity sensors based on hydrogenated
  amorphous carbon films}.
\newblock \bibinfo{journal}{Materials Research Express} \bibinfo{volume}{6},
  \bibinfo{pages}{025604}.
\bibitem[{Ferrari and Basko(2013)}]{ferrari2013raman}
\bibinfo{author}{Ferrari, A.C.}, \bibinfo{author}{Basko, D.M.},
  \bibinfo{year}{2013}.
\newblock \bibinfo{title}{Raman spectroscopy as a versatile tool for studying
  the properties of graphene}.
\newblock \bibinfo{journal}{Nature nanotechnology} \bibinfo{volume}{8},
  \bibinfo{pages}{235--246}.
\bibitem[{Ferrari and Robertson(2000)}]{ferrari2000interpretation}
\bibinfo{author}{Ferrari, A.C.}, \bibinfo{author}{Robertson, J.},
  \bibinfo{year}{2000}.
\newblock \bibinfo{title}{Interpretation of raman spectra of disordered and
  amorphous carbon}.
\newblock \bibinfo{journal}{Physical Review B} \bibinfo{volume}{61},
  \bibinfo{pages}{14095}.
\bibitem[{Ferrari and Robertson(2004)}]{ferrari2004raman}
\bibinfo{author}{Ferrari, A.C.}, \bibinfo{author}{Robertson, J.},
  \bibinfo{year}{2004}.
\newblock \bibinfo{title}{Raman spectroscopy of amorphous, nanostructured,
  diamond--like carbon, and nanodiamond}.
\newblock \bibinfo{journal}{Philosophical Transactions of the Royal Society of
  London A: Mathematical, Physical and Engineering Sciences}
  \bibinfo{volume}{362}, \bibinfo{pages}{2477--2512}.
\bibitem[{Frattini et~al.(2021)Frattini, Torres, Silva, Repetto, Gomez and
  Dobry}]{frattini2021effect}
\bibinfo{author}{Frattini, G.}, \bibinfo{author}{Torres, S.},
  \bibinfo{author}{Silva, L.}, \bibinfo{author}{Repetto, C.E.},
  \bibinfo{author}{Gomez, B.J.A.J.}, \bibinfo{author}{Dobry, A.},
  \bibinfo{year}{2021}.
\newblock \bibinfo{title}{The effect of nitriding on the humidity sensing
  properties of hydrogenated amorphous carbon films}.
\newblock \bibinfo{journal}{Physica Scripta} .
\bibitem[{Jarosi\'nski et~al.(2019)Jarosi\'nski, Pawlak and
  Al-Ani}]{Jarosinski2019inverse}
\bibinfo{author}{Jarosi\'nski, L.}, \bibinfo{author}{Pawlak, J.},
  \bibinfo{author}{Al-Ani, S.K.J.}, \bibinfo{year}{2019}.
\newblock \bibinfo{title}{Inverse logarithmic derivative method for determining
  the energy gap and the type of electron transitions as an alternative to the
  tauc method}.
\newblock \bibinfo{journal}{Optical Materials} \bibinfo{volume}{88},
  \bibinfo{pages}{667--673}.
\bibitem[{Muñoz and Gómez-Aleixandre(2013)}]{Munoz2013}
\bibinfo{author}{Muñoz, R.}, \bibinfo{author}{Gómez-Aleixandre, C.},
  \bibinfo{year}{2013}.
\newblock \bibinfo{title}{Review of cvd synthesis of graphene}.
\newblock \bibinfo{journal}{Chemical Vapor Deposition} \bibinfo{volume}{19},
  \bibinfo{pages}{297--322}.
\newblock \URLprefix
  \url{https://onlinelibrary.wiley.com/doi/abs/10.1002/cvde.201300051},
  \DOIprefix\doi{https://doi.org/10.1002/cvde.201300051}.
\bibitem[{Robertson(1993)}]{Robertson1993}
\bibinfo{author}{Robertson, J.}, \bibinfo{year}{1993}.
\newblock \bibinfo{title}{Deposition of diamond-like carbon}.
\newblock \bibinfo{journal}{Philosophical Transactions of the Royal Society of
  London. Series A: Physical and Engineering Sciences} \bibinfo{volume}{88},
  \bibinfo{pages}{277–286}.
\bibitem[{Sadezky et~al.(2005)Sadezky, Muckenhuber, Grothe, Niessner and
  P{\"o}schl}]{sadezky2005raman}
\bibinfo{author}{Sadezky, A.}, \bibinfo{author}{Muckenhuber, H.},
  \bibinfo{author}{Grothe, H.}, \bibinfo{author}{Niessner, R.},
  \bibinfo{author}{P{\"o}schl, U.}, \bibinfo{year}{2005}.
\newblock \bibinfo{title}{Raman microspectroscopy of soot and related
  carbonaceous materials: Spectral analysis and structural information}.
\newblock \bibinfo{journal}{Carbon} \bibinfo{volume}{43},
  \bibinfo{pages}{1731--1742}.
\bibitem[{Tai et~al.(2009)Tai, Lee, Chen, Wei and Chang}]{Tai2009}
\bibinfo{author}{Tai, F.C.}, \bibinfo{author}{Lee, S.C.},
  \bibinfo{author}{Chen, J.}, \bibinfo{author}{Wei, C.},
  \bibinfo{author}{Chang, S.H.}, \bibinfo{year}{2009}.
\newblock \bibinfo{title}{Multipeak fitting analysis of raman spectra on dlch
  film}.
\newblock \bibinfo{journal}{J. Raman Spectrosc.} \bibinfo{volume}{40},
  \bibinfo{pages}{1055--1059}.
\bibitem[{Tauc(1968)}]{tauc1968optical}
\bibinfo{author}{Tauc, J.}, \bibinfo{year}{1968}.
\newblock \bibinfo{title}{Optical properties and electronic structure of
  amorphous ge and si}.
\newblock \bibinfo{journal}{Materials research bulletin} \bibinfo{volume}{3},
  \bibinfo{pages}{37--46}.
\bibitem[{Wang et~al.(2020)Wang, Ren, Hou, Yan, Liu, Zhang, Zhang and
  Guo}]{wang2020}
\bibinfo{author}{Wang, J.b.}, \bibinfo{author}{Ren, Z.}, \bibinfo{author}{Hou,
  Y.}, \bibinfo{author}{Yan, X.l.}, \bibinfo{author}{Liu, P.z.},
  \bibinfo{author}{Zhang, H.}, \bibinfo{author}{Zhang, H.x.},
  \bibinfo{author}{Guo, J.j.}, \bibinfo{year}{2020}.
\newblock \bibinfo{title}{A review of graphene synthesisatlow temperatures by
  cvd methods}.
\newblock \bibinfo{journal}{New Carbon Materials} \bibinfo{volume}{35},
  \bibinfo{pages}{193--208}.
\bibitem[{Wu et~al.(2018)Wu, Lin, Cong, Liu and Tan}]{Wu2018}
\bibinfo{author}{Wu, J.B.}, \bibinfo{author}{Lin, M.L.}, \bibinfo{author}{Cong,
  X.}, \bibinfo{author}{Liu, H.N.}, \bibinfo{author}{Tan, P.H.},
  \bibinfo{year}{2018}.
\newblock \bibinfo{title}{Raman spectroscopy of graphene-based materials and
  its applications in related devices}.
\newblock \bibinfo{journal}{Chem. Soc. Rev.} \bibinfo{volume}{47},
  \bibinfo{pages}{1822--1873}.
\newblock \URLprefix \url{http://dx.doi.org/10.1039/C6CS00915H},
  \DOIprefix\doi{10.1039/C6CS00915H}.
\bibitem[{Yuan and Mayanovic(2017)}]{Yuan2017}
\bibinfo{author}{Yuan, X.}, \bibinfo{author}{Mayanovic, R.},
  \bibinfo{year}{2017}.
\newblock \bibinfo{title}{An empirical study on raman peak fitting and its
  application to raman quantitative research}.
\newblock \bibinfo{journal}{Applied Spectroscopy} \bibinfo{volume}{71},
  \bibinfo{pages}{2325--2338}.

\end{thebibliography}
\end{document}